\documentstyle[pra,aps]{revtex} 
\begin{document}

\title{DECOHERENCE CONTROL IN QUANTUM INFORMATION PROCESSING: 
SIMPLE MODELS}

\author{Lorenza Viola and Seth Lloyd\thanks{Electronic addresses: 
vlorenza@mit.edu; slloyd@mit.edu} }
\address{d'Arbeloff Laboratory for Information Systems and Technology, \\
Department of Mechanical Engineering,  Massachusetts Institute of 
Technology, \\ Cambridge, Massachusetts 02139 } 

\maketitle

\begin{abstract}
{We explore a strategy for protecting the evolution of a qubit 
against the effects of environmental noise based on the application of 
controlled time-dependent perturbations. In the case of a purely decohering 
coupling, an explicit sequence of control operations is designed, able to 
average out the decoherence of the qubit with high efficiency. We argue that, 
in principle, the effects of arbitrary qubit-environment interactions can be 
removed through suitable decoupling perturbations acting on the system 
dynamics over time scales comparable to the correlation time of the 
environment.}
\end{abstract}

\section{INTRODUCTION}

Decoherence remains one of the most serious obstacles to the exploitation 
of the speed-up promised by quantum computation {\cite{review}}. 
Broadly speaking, two different philosophies are being investigated to 
overcome the decoherence problem. On one hand, {\sl passive} error-prevention 
schemes have been proposed, based on the idea of encoding logical quantum bits 
(qubits) within subspaces which do not decohere owing to symmetry properties
{\cite{rasetti,lidar}}. 
On the other hand, {\sl active} error-correction 
approaches have been formalized within a sophisticated theory of quantum 
error-correcting codes (QECC), where a logical qubit is encoded in the 
larger Hilbert space of several physical qubits and suitable feedback 
operations are conditionally carried out {\cite{shor,nmr1}}. 

Although a purposeful manipulation is implied in the latter case, quantum 
error-correcting codes can be properly interpreted in terms of a clever 
redundancy in
the software architecture rather than a physical way to operate on 
decoherence. In this work, we explore the possibility of using control 
techniques to modify and eliminate decoherence. Unlike recent proposals for
feedback (or closed-loop) control schemes of decoherence in quantum optical 
systems {\cite{vitali}}, we apply control in the simpler open-loop 
configuration to general models of quantum information processing
systems {\cite{viola}}. 
The underlying idea is suggested by high-resolution
pulsed Nuclear Magnetic Resonance (NMR), where astonishingly versatile 
refocusing and decoupling techniques are nowadays available to remove the 
effects of interactions among the spins that are considered unwanted or 
uninteresting {\cite{ernst}}. 
In our analysis, we outline the conditions under
which analogous procedures can be extended from eliminating interactions 
{\sl internal} to the system to suppressing interactions of the system with 
an {\sl external} quantized environment. In particular, the role of the 
environment correlation time will be pointed out as a further parameter to be
engineered in the struggle for preserving quantum coherence. 
    
\section{QUANTUM BANG-BANG CONTROL OF QUBIT DECOHERENCE} 

We start by investigating a prototype situation that conveys the basic idea
in the simplest form. We will focus on the dynamics of a single memory cell
of quantum information (qubit) undergoing decoherence due to the coupling
to a thermal reservoir. The physical qubit can be associated either to a 
fictitious or to a real spin-$1/2$ system, the latter case allowing for a 
direct reference to the language of Nuclear Magnetic Resonance and NMR 
quantum computation {\cite{ernst,nmrqc}}. 
We assume that the fastest relaxation
process originated by the interaction with the quantized environment is a
{\sl purely dephasing} process i.e., in NMR terminology, 
no dissipative $T_1$-type of decay takes place. 
A general model for the dynamics of the overall qubit 
$+$ bath system is then provided by the purely decohering spin-boson 
Hamiltonian ($\hbar=1$):
\begin{equation}
H_0 = H_S + H_B +H_{SB} = {\omega_0 \over 2} \, \sigma_z + 
\sum_k \,  \omega_k b_k^\dagger b_k + 
\sigma_z \sum_k \, (g_k b_k^\dagger + g_k^\ast b_k) \;.    
\label{hamiltonian}
\end{equation}
Here, $\sigma_z$ is the standard diagonal Pauli matrix, with qubit basis 
states denoted as $|i \rangle$, $i=0,1$, while $b_k^\dagger,b_k,g_k$ are 
bosonic operators and coupling parameters for the $k$-th field mode
respectively. Hamiltonian (\ref{hamiltonian}) is widely used in the quantum 
computation literature to investigate the effect of phase errors, 
representing the most nonclassical and dangerous source of errors for quantum 
qubits {\cite{unruh,palma,ban}}. 
Since $[\sigma_z, H_0]=0$, spin populations
are not affected by time evolution and decoherence dynamics is characterized 
completely by the qubit coherence with respect to the computational 
basis:
\begin{equation}
\rho_{01}(t) \equiv \langle 0| \, \rho_S (t) \, | 1 \rangle = 
\langle 0| \, \mbox{Tr}_B \{ \rho_{tot}(t) \} \, | 1 \rangle = 
\mbox{Tr}_{B,S}\{ \rho_{tot}(t_0)\sigma_+(t) \} \equiv 
\langle \sigma_+ (t) \rangle \;,
\label{coherence}
\end{equation}
where, starting from the left and using standard notations, the relevant
reduced density matrix element $\rho_{01}(t)=\rho_{10}^\ast(t)$ in the 
Schr\"odinger picture is linked to the expectation value of the time-evolved 
ladder operator in the Heisenberg representation  
\begin{equation}
\sigma_+(t) =  {1 \over 2} ( \sigma_x(t) + i\, \sigma_y(t) )\;.
\end{equation}

The evolution of the qubit coherence (\ref{coherence}) can be calculated
exactly under the customary assumptions about the initial state of the overall
system, i.e. qubit and environment are initially uncorrelated and the 
environment is in thermal equilibrium at a temperature $T$. In the Heisenberg
representation, one may write formally 
\begin{equation}
\langle \sigma_+(t) \rangle  =  \langle G_{tot}(t_0,t)\, 
\sigma_+(t_0) \rangle \;,   \label{prop}
\end{equation}
with the propagator $G_{tot}(t_0,t)$ determined by the solution of the 
Heisenberg equations for the coupled spin $+$ bath motion:
\begin{equation}
\left\{ \begin{array}{lcl}
\dot{\sigma}_z (t) & = & 0 \;, \\
\dot{\sigma}_+ (t) & = & i \omega_0 \sigma_+ (t) + 2i \sum_k \, 
             (g_k b_k^\dagger (t) + g_k^\ast b_k (t) )\, \sigma_+ (t)\;, \\
\dot{b}_k (t) & = & -i \omega_k b_k (t) -i g_k \sigma_z (t) \;, \\
\dot{b}_k^\dagger (t) & = & +i \omega_k b^\dagger_k (t) +
i g_k^\ast \sigma_z (t) \;. \\
\end{array} \right. 
\label{eqsmotion}
\end{equation}
The result can be written in the following form {\cite{viola}}:
\begin{equation}
\rho_{01}(t)=\mbox{e}^{i \omega_0 (t-t_0) - \Gamma_0(t-t_0)} \, 
\rho_{01}(t_0) \;, 
\label{rhodit}
\end{equation}
the loss of phase information being characterized by the damping function 
\begin{equation}
\Gamma_0 (t-t_0) \equiv \sum_k \Gamma_0 (k; t-t_0) =
\sum_k 4 |g_k|^2 \coth \bigg({\omega_k \over 2T} \bigg)
\, {1-\cos \omega_k (t-t_0) \over \omega_k^2} \;,  
\label{gamma0}
\end{equation}
in units where the Boltzmann constant $k_B=1$. In the limit of a truly 
macroscopic environment, a description in terms of a continuum of modes is
appropriate and the dependence of the decoherence function (\ref{gamma0})
on reservoir properties can be cast in a compact form after introducing the
spectral density function $I(\omega)$, 
\begin{equation}
\Gamma_0 (t-t_0) \equiv \int_0^\infty d\omega \, \Gamma_0(\omega; t-t_0) =
\int_0^\infty d\omega \, I(\omega) \,
4 \coth\bigg( {\omega \over 2T} \bigg) 
{1-\cos \omega(t-t_0) \over \omega^2} \;.
\label{spectral}
\end{equation}
Depending on the temperature $T$ and the spectral density $I(\omega)$, 
qualitatively different open-system evolutions arise in general, with a 
different interplay between quantum fluctuation and dissipation phenomena.
Regardless the details of the spectral density function, however, the 
existence of a certain ultraviolet cut-off frequency $\omega_c$ is always
demanded on physical grounds, leading to   
\begin{equation}
I(\omega) \rightarrow 0 \hspace{2cm}\mbox{for  } \omega > \omega_c\;. 
\end{equation}
%\newpage
Although the specific value of $\omega_c$ depends on a natural cut-off 
frequency varying from system to system, $\omega_c$ can be generally 
associated to a characteristic time $\tau_c \sim \omega_c^{-1}$ setting the
fastest (finite !) time scale of the irreversible dynamics. $\tau_c$ is 
known as the {\sl correlation time} of the environment. The dynamics of the 
decoherence process arising from (\ref{spectral}) for various choices of 
$I(\omega)$ has been investigated in detail 
elsewhere {\cite{viola,palma}}.
A pictorial representation for the important class of Ohmic reservoirs,  
$I(\omega) \propto \omega \, \mbox{e}^{-\omega/\omega_c}$, is shown in Fig. 1.

We introduce now a procedure aimed at improving the coherence properties of 
the qubit by adding a controllable time-dependent interaction to the original
Hamiltonian:
\begin{equation}
H(t)=H_0 + H_1(t) = H_S + H_B + H_{SB} + H_1(t) \;. 
\label{ham2}
\end{equation}
In the same spirit underlying multiple-pulse techniques in the manipulation
of nuclear spin Hamiltonians {\cite{ernst}}, we try to average out the 
unwanted effects of the qubit-reservoir coupling $H_{SB}$ by applying a 
sequence of coherent $\pi$-pulses that repetitively flip the state of the 
system. Under the assumptions that the duration of the pulses is short 
enough compared to the typical decoherence time and the strength of the 
control field is sufficient to override the $H_{SB}$ coupling, the 
pulsed-mode operation allows one to separate the actions of the bath 
and the
external field, by neglecting $H_{SB}$ while $H_1(t)$ is on. Specifically, 
$H_1(t)$ is assumed to schematize a train of $n_P$ identical $\pi$-pulses 
along the $\hat{x}$-axis applied on resonance at instants $t=t^{(n)}_P$,
$n=1,\ldots,n_P$, with pulse separation 
$t^{(n+1)}_P-t^{(n)}_P = \Delta t$. By invoking, as usual, the rotating-wave
approximation, we have 
\begin{equation}
H_1(t)= \Xi(t)\, [\,\cos( \omega_0 t ) \sigma_x + 
\sin  ( \omega_0 t) \sigma_y \,] 
\;, \label{rf}
\end{equation}
with envelope function 
\begin{equation}
\Xi (t) = \sum_{n=1}^{n_P} \, V \,\Big[ \vartheta (t- t^{(n)}_P) - 
\vartheta (t- t^{(n)}_P - \tau_P) \Big] \;, \hspace{0.5cm}
t^{(n)}_P = t_0 + n \Delta t, \; n=1,\ldots,n_P\;. 
\label{envelope}
\end{equation}
In Eq. (\ref{envelope}), $\vartheta(\cdot)$ denotes the Heaviside 
step function,  
and the height $V$ and the width $\tau_P$ of each pulse satisfy 
$2V \tau_P = \pi$. To simplify things, we work henceforth in the limit of 
infinitely narrow pulses $\tau_P \rightarrow 0$, assuming the kicks of 
radiofrequency control field large enough to produce instantaneous spin 
rotations. By 
analogy with the classical technique of {bang-bang} (or on-off) controls, 
whereby piecewise controls with extremal values are 
exploited {\cite{bang}}, 
one may look at this strategy as an implementation of {\sl quantum 
bang-bang control}. 

In order to depict the evolution associated to a given pulse sequence, it is 
convenient to think the latter as formed by repeated elementary cycles of 
spin-flips, a complete cycle being able to return the spin back to the 
starting configuration. For definiteness, let us analyze the first cycle, 
made of the following steps: evolution under $H_0$ during 
$t_0 \leq t \leq t^{(1)}_P$; $\pi$-pulse $P_1$ at time $t^{(1)}_P$; 
evolution under $H_0$ during $t_P^{(1)} \leq t \leq t^{(2)}_P$; $\pi$-pulse 
$P_2$ at time $t^{(2)}_P$. After a total time $t_1=t_0 + 2 \Delta t$, the first
cycle is complete. The description of $\pi$-pulses turns out to be extremely 
simple in the Heisenberg representation. Nothing happens to the bath operators
$b_k,b^\dagger_k$ in the limit of instantaneous pulses, while, by denoting 
with $t_P^{-\,(+)}$ the times immediately before (after) a pulse respectively, 
spin operators are transformed as follows:
\begin{equation}
\left\{ \begin{array}{lcl}
 \sigma_z(t^+_P) & = & -\, \sigma_z(t^-_P)\;, \\
 \sigma_+(t_P^+) & = & [\sigma_+(t_P^-)]^\dagger\;. 
\end{array} \right.
\label{a3}
\end{equation}
\noindent
In terms of the free propagator $G_{tot}(t_i,t_j)$ introduced in (\ref{prop}) 
to evolve coherence from $t_i$ to $t_j$, the time development during the 
cycle can be represented as  
\begin{equation}
\langle \sigma_+(t_0 + 2 \Delta t) \rangle = \langle G_{tot}(t_0,t_0+\Delta t) \,
\sigma_+(t_0) \, 
G_{tot}^\dagger (t_0+ \Delta t, t_0+ 2 \Delta t) \rangle \;, 
\label{a5}
\end{equation}
to be compared with 
\begin{equation}
\langle \sigma_+(t_0 + 2 \Delta t) \rangle = \langle 
G_{tot}(t_0+\Delta t, t_0+ 2\Delta t) \,G_{tot}(t_0, t_0+ \Delta t)\, 
\sigma_+(t_0) \rangle  
\label{a6}
\end{equation}
in the absence of pulses. Since instantaneous rotations introduce 
discontinuous 
changes in operators (\ref{a3}), care must be taken in evaluating the two
propagators $G_{tot}(t_0,t_0+\Delta t)$, $G_{tot}(t_0+\Delta t,t_0+2 \Delta t)$
separately, by solving Heisenberg equations of motion (\ref{eqsmotion}) with 
initial conditions at $t=t_0$, $t=t_P^+=t_0 + \Delta t$ respectively. Only at 
the end of the calculation everything can be expressed with respect to the 
initial time of the cycle. Omitting the details, the result for the coherence 
evolution over the first complete cycle is {\cite{viola}} 
\begin{equation}
\rho_{01}(t_0+2 \Delta t)=\mbox{e}^{ - \Gamma_P(N=1,\Delta t)} \, 
\rho_{01}(t_0) \;, 
\label{rhopdit}
\end{equation}
where a new decoherence function for $N=1$ spin cycles has been introduced:
\begin{equation}
\Gamma_P (N=1,\Delta t) = \sum_k \Gamma_0 (k; 2 \Delta t) \, 
\bigg| 1 - 2 {1- \mbox{e}^{i \omega_k \Delta t} \over 
                1- \mbox{e}^{2i \omega_k \Delta t} } \bigg|^2   \;.  
\label{gammap1}
\end{equation}
Since, for each mode, the additional factor arising from the pulses is of 
order $O(\omega_k^2 \Delta t^2)$ $\ll 1$ for small $\Delta t$, we may guess 
that something interesting is happening in a regime where the state of the
qubit is tipped very rapidly. This is made clear by generalizing the 
description to an arbitrary number $N$ of spin-flip cycles, involving a total
number of $n_P=2N$ $\pi$-pulses. After straight forward calculations along the
same line outlined above, the expression for the qubit coherence at the 
final time $t_N=t_0 + 2N \Delta t$ is the following:
\begin{equation}
\rho_{01}(t_0+2 N \Delta t)=\mbox{e}^{ - \Gamma_P(N,\Delta t)} \, 
\rho_{01}(t_0) \;, 
\label{rhopndit}
\end{equation}
with 
\begin{eqnarray}
\Gamma_P (N,\Delta t) & = & \sum_k \Gamma_0 (k; 2 N \Delta t) \, 
\Big| 1- f_k(N,\Delta t) \Big|^2 \;, \label{gammap} \\
f_k(N,\Delta t) & = & 2 \, {1- \mbox{e}^{i \omega_k \Delta t} \over 
                1- \mbox{e}^{2i \omega_k \Delta t} } \,
\sum_{n=1}^N \, \mbox{e}^{2i(n-1) \omega_k \Delta t}  \;.  \label{fk}
\end{eqnarray}
From inspection of Eq. (\ref{gammap}), the contribution due to the pulse 
sequence turns out to manifest in the typical form of an interference factor. 
The implications for the decoherence properties are easily stated by 
considering the mathematical limit where 
$\Delta t \rightarrow 0$, $N \rightarrow \infty$, subjected to the constraint 
$2N\Delta t = t_N-t_0$. Under these conditions, one can prove 
that {\cite{viola}}
\begin{equation}
\lim_{\Delta t \rightarrow 0} f_k(N, \Delta t)  = 1 \hspace{1cm} \forall k 
\hspace {1cm} \Rightarrow \hspace{1cm}\lim_{\Delta t \rightarrow 0} 
\Gamma_P(N, \Delta t) =0 \;.  
\label{suppress}
\end{equation}
Thus, in the limit of continuous flipping, decoherence is completely washed
out for any temperature and any spectral density function.  

Obviously, a continuous limit of this kind is scarcely meaningful from a 
physical point of view. However, this ideal situation should be approached 
if $\Delta t$ is made small compared to the fastest characteristic time 
operating within the environmental noise, i.e. the reservoir correlation 
time $\tau_c$. Hence, we expect that a sufficient condition in order to meet 
(\ref{suppress}) and suppress decoherence is   
\begin{equation}
\Delta t \ll \tau_c \;. 
\label{condition}
\end{equation}
More explicitly, this result implies that, given an arbitrary time 
$t$, one can always recover the initial state and the coherence of the qubit 
by making $t$ the end time of a pulse sequence and by adjusting the parameters
to satisfy $t=t_N=2N \Delta t$ and $\Delta t \ll \tau_c$. Then, at time
$t$, a {\sl coherence echo} is formed. Alternatively, by keeping the qubit 
flipped and restricting the observation to cycle times $t_N$, $N=1,2,\ldots$, 
the system is found to evolve ideally as it would do in the absence of the 
coupling $H_{SB}$ responsible for decoherence.
A typical behavior originated by the pulsing procedure for the prototype
high-temperature Ohmic environment of Fig. 1 is displayed in Fig. 2.

So far, the suppression of decoherence has been derived in a rather formal 
way. Actually, a simple physical explanation can be provided as well. 
Similarly to the original well-known spin-echo 
phenomenon {\cite{hahn}}, and
to the more sophisticated solid-echoes or magic echoes 
experiments {\cite{ernst}}, 
the basic argument here is a time-reversal argument. 
The examination of a single elementary spin-flip cycle suffices to capture 
the underlying mechanism. Roughly speaking, and looking back at the 
representation (\ref{prop}), it is the presence of the transformed propagator 
$G_{tot}^{-1}(t_0+\Delta t, t_0 + 2 \Delta t)$, generated by the couple 
of $\pi$-pulses,  that {\sl simulates} the effect of a time-reversal. Would 
the evolution during the second half of the cycle be identical to the one in 
the first $\Delta t$ interval, then it would be  
\begin{equation}
G_{tot}^{-1}(t_0+\Delta t, t_0 + 2 \Delta t)=
G_{tot}^{-1}(t_0,t_0+\Delta t)\;, 
\end{equation} 
and, therefore, 
$\langle \sigma_+(t_0 + 2 \Delta t) \rangle =   
\langle \sigma_+(t_0) \rangle$ as a consequence of the cyclic property in the 
trace. Instead, this reversal is only approximate in general since the two
propagations differ by a dephasing factor $\mbox{e}^{i \omega_k \Delta t}$ in 
the evolution of each reservoir mode {\cite{viola}}. 
However, if the condition 
(\ref{condition}) is met, then the cycle is effectively equivalent to an 
exact time-reversal and, by iteration on every cycle, the elimination of 
decoherence (\ref{suppress}) is achieved. 
  
\section{DYNAMICAL DECOUPLING OF QUBIT-ENVIRONMENT INTERACTIONS}

In this section, we rederive the result established above in a form that 
opens up the way to further generalization. The first step is to formally 
reinterpret the method of reducing environment-induced decoherence by 
successive application of 
$\pi$-pulses within the general framework of decoupling techniques based on 
{\sl controlled averaging}. In NMR, sophisticated decoupling schemes are 
routinely used to simplify complex spectra by manipulating the underlying 
spin Hamiltonian to an extent allowing for a successful 
analysis {\cite{ernst,ulrich}}.
In particular, a relevant class of decoupling procedures, including 
spin-decoupling and multiple-pulse experiments, involves selective averaging 
in the internal spin space. The idea is to introduce controlled motions into
the system, with the time-dependence designed in such a way that undesired 
terms in the Hamiltonian are averaged out. In extending similar techniques to
the decoupling of interactions between a system and its environment, the major 
difference stems from the fact that the decoupling action can be easily 
exerted only on the system variables, the bath degrees of freedom being 
generally uncontrollable. 

Looking back at the Hamiltonian (\ref{hamiltonian}), we start by 
seeking a perturbation $H_1(t)$ to be added as a suitable decoupling 
interaction in order to remove $H_{SB}$, Eq. (\ref{ham2}). We restrict to a 
situation where $H_1(t)$ is {\sl cyclic}, i.e. satisfying the following 
conditions ($t_0=0$ henceforth):
\begin{eqnarray}
 & (i) \hspace{2mm}&  H_1(t)=H_1(t+ \Delta t) 
\hspace{1cm}\mbox{for some }\Delta t\; ; \label{cycle1} \\
 & (ii) \hspace{2mm}&  U_1(t)\equiv T\exp\bigg\{\hspace{-1mm} 
-i \int_0^t\; ds \,H_1(s) \bigg\} = U_1(t+T_c)\hspace{1cm}\mbox{for some }
T_c\;.   \label{cycle2}
\end{eqnarray}
From $(ii)$, $U_1(T_c)=${\sf I} and $T_c$ is called the {\sl cycle time}.  
An elegant description of the dynamics arising in the presence of a cyclic 
perturbation is provided by the so-called average Hamiltonian 
theory {\cite{ernst,ulrich}}. 
We only recall here the basic ingredients in a language 
that is appropriate to quantum information processing. In particular, we
schematize the perturbation as a sequence on $n_P$ ideal pulses (or logic
gates), represented by unitary transformations $P_1, \ldots, P_{n_P}$, and 
separated by free evolution periods under $H_0$. 
The operators $\{ P_i \}$ fulfill
$U_1(T_c)=P_{n_P}\ldots P_2 P_1 =${\sf I} by cyclicity. Average Hamiltonian
theory is based on moving to the time-dependent interaction representation
with respect to $H_1(t)$ defined, as usual, by 
\begin{equation}
\rho_{tot}(t)=U_1(t) \tilde{\rho}_{tot}(t) U_1^\dagger (t) \;, 
\label{toggling}
\end{equation}
with
\begin{equation}
{d \over dt} \tilde{\rho}_{tot}(t) = -i \Big[ \tilde{H}(t), 
\tilde{\rho}_{tot}(t) \Big]\;,
\hspace{2cm} \tilde{H}(t) = U_1^\dagger (t) H_0 U_1(t) \;.    
\label{htilde}
\end{equation}
In (\ref{htilde}), $H_0$ is given by (\ref{hamiltonian}) and, according to 
the standard NMR literature, the transformed Hamiltonian $\tilde{H}(t)$ is 
also known as the {\sl toggling frame Hamiltonian}. If $U_{tot}(t)$ and 
$\tilde{U}_{tot}(t)$ denote the time evolution operators in the Schr\"odinger 
and interaction picture respectively, due to (\ref{cycle2}) one gets
\begin{equation}
U_{tot}(nT_c)=U_1(nT_c) \tilde{U}_{tot}(nT_c)= \tilde{U}_{tot}(nT_c)
=\Big[ \tilde{U}_{tot}(T_c)\Big]^n \equiv \mbox{e}^{-i \overline{H} t_n} \;, 
\label{average}
\end{equation}
i.e. provided the observation of the dynamics is restricted to {\sl 
stroboscopic} and {\sl synchronized} sampling, $t_n=nT_c$, it is 
sufficient to know the evolution over a single cycle as described by the 
transformed Hamiltonian $\tilde{H}(t)$. In (\ref{average}), the last equality
defines the average Hamiltonian and contains the main result of average 
Hamiltonian theory: for suitable repeated observations, the motion of the 
system under the influence of the time-dependent field $H_1(t)$ can be 
represented by a {\sl constant} average Hamiltonian $\overline{H}$. The 
calculation of $\overline{H}$ is usually performed on the basis of a standard 
Magnus expansion of the time-ordered exponential defining 
$\tilde{U}_{tot}(T_c)$ {\cite{ulrich}}, i.e.
\begin{equation}
\tilde{U}_{tot}(T_c) = T\exp \bigg\{ -i \int_0^{T_c} \, ds \, \tilde{H}(s)
\bigg\} \equiv \mbox{e}^{-i \overline{H} T_c} = 
\mbox{e}^{-i[\, \overline{H}^{(0)} + \overline{H}^{(1)} + \ldots\,] T_c} \;. 
\label{magnus}
\end{equation}
In our case, the evaluation of (\ref{magnus}) is simplified since the toggling
frame Hamiltonian $\tilde{H}(t)$ is piecewise constant during the intervals
$\Delta t_k$ separating consecutive rotations and one can introduce stepwise
transformed Hamiltonians:
\begin{equation}
\tilde{U}_{tot}(T_c)= \mbox{e}^{-i \tilde{H}_{n_P} \Delta t_{n_P} }
\ldots \mbox{e}^{-i \tilde{H}_0 \Delta t_0 }\;, \hspace{2cm}
\tilde{H}_k= (P_k \ldots P_1)^{-1} \,H_0\,(P_k\ldots P_1) \;, 
\label{stepwise}
\end{equation} 
where, obviously, $\sum_k \Delta t_k=T_c$. Thus, the lowest-order 
approximation $\overline{H}^{(0)}$ to the average Hamiltonian in 
(\ref{magnus}) has a particularly simple form:
\begin{equation}
\overline{H}^{(0)}= {1 \over T_c} \bigg\{ \tilde{H}_0 \Delta t_0 +
\ldots + \tilde{H}_{n_P} \Delta t_{n_P} \bigg\} = {1 \over T_c}
\sum_{k=0}^{n_P}\, \Delta t_k \, P_1^{-1} \ldots P_k^{-1} \,H_0\,
P_k \ldots P_1 \;. \label{zeroth}
\end{equation}
In addition, explicit expressions for the corrections arising from 
higher-order terms are systematically available from the Magnus 
expansion {\cite{ulrich}}.
Loosely speaking, the goal of decoupling is to devise a tranformation to a 
toggling frame (\ref{toggling}), where the unwanted coupling $H_{SB}$ no longer
appears up to a certain order $\overline{H}^{(r)}$ and higher-order 
contributions $\overline{H}^{(r+1)},\ldots$ are made neglegible. 

We are now in a position to apply this formalism to the evolution of the 
decohering qubit, whose Hamiltonian we rewrite for convenience as follows:
\begin{equation}
H_0 =H_S +H_B + H_{SB} = {\omega_0 \over 2} \sigma_z +\sum_k \,\omega_k
b_k^\dagger b_k + \sigma_z {\cal B}_z\;, \hspace{1cm} 
{\cal B}_z =\sum_k \, (g_k b_k^\dagger +g_k^\ast b_k) \;. 
\label{ham3}
\end{equation}
Using (\ref{htilde}), the transformation to the toggling frame associated to 
$H_1(t)$ leads to
\begin{equation}
\tilde{H}(t)=H_B + U_1^\dagger (t) \,H_S \,U_1(t) + [ U_1^\dagger (t) \,
\sigma_z \, U_1(t) ] \, {\cal B}_z \;, 
\label{tilde2}
\end{equation}
and to a zero-th order average Hamiltonian given by (\ref{zeroth}):
\begin{equation}
\overline{H}^{(0)}= H_B + {1 \over T_c} \sum_{k=0}^{n_P}\, \Delta t_k 
{\cal P}_k^{-1} H_S {\cal P}_k + \bigg[ {1 \over T_c} \sum_{k=0}^{n_P}\, 
\Delta t_k {\cal P}_k^{-1} \sigma_z {\cal P}_k \bigg] {\cal B}_z \;,
\label{zero}
\end{equation}
where, in (\ref{tilde2}) and (\ref{zero}), the fact that reservoir operators 
are unaffected by the control field has been evidenced and the short 
notation ${\cal P}_k=P_k \ldots P_1$ has been introduced. The second and 
third terms in (\ref{zero}) correspond, in general, to a transformed qubit
Hamiltonian and a transformed qubit-bath interaction. It is immediate to 
realize that the effect of the $\pi$-pulse sequence of Sec. II viewed in this
frame is to cause the latter interaction term to vanish. To make the 
identification explicit, we rearrange the elementary spin-flip cycle as 
follows:
\begin{equation}
 - \; \underbrace{ {\Delta t \over 2} \; - \; P_x^{180}\; -\; 
{\Delta t \over 2}\; -\; 
{\Delta t \over 2}\; -\; P_x^{180}\; -\; {\Delta t \over 2} }_{cycle} \; -\;
{\Delta t \over 2}\; \ldots 
\label{cp}
\end{equation}
where the rotation axis of the pulses has been indicated and, to compare with 
(\ref{zeroth}), $\Delta t_0=\Delta t_2=\Delta t/2$, $\Delta t_1=\Delta t$,
$T_c=2\Delta t$. Written in the form (\ref{cp}), the decoupling sequence for
(\ref{ham3}) is nothing but a variant of the famous {\sl Carr-Purcell} (CP) 
sequence, that is ordinarily exploited to get rid of static applied-field
inhomogeneities {\cite{ernst,ulrich}}. 
At variance with this standard usage of
the CP-sequence, however, where the size of $\Delta t$ is of no importance,
the averaging of $H_{SB}$ at zeroth-order does not guarantee by itself the 
elimination of decoherence. Obviously, higher-order terms in (\ref{magnus})
have to be quenched. The question under which circumstances the Magnus series
can be truncated after the leading term or the few lowest-order corrections
is nontrivial.
It is possible to show {\cite{no2}} that, 
since the $r$-th order contribution 
$\overline{H}^{(r)}=O(\Delta t^r)$, a sufficient condition is $\Delta t 
\, \omega_c \ll 1$, which is identical to (\ref{condition}). 
Then, from (\ref{average}), one ideally gets the stroboscopic equality
\begin{equation}
\rho_S(nT_c)= \mbox{Tr}_B \{ \, \mbox{e}^{-i \overline{H}^{(0)} n T_c}
\, \rho_S(0) \rho_B(0) \, \mbox{e}^{i \overline{H}^{(0)} n T_c} \,\}
= \rho_S(0) \;.
\end{equation}

The generalization of the above scheme to decouple arbitrary
qubit-environment interactions is, in principle, straight forward since the 
most general bilinear coupling can be expressed as a mixed sum of 
{\sl error-generators} {\cite{rasetti,knill}}:
\begin{equation}
H_{SB}=\sigma_x {\cal B}_x +\sigma_y {\cal B}_y +\sigma_z {\cal B}_z \;,
\label{arbitrary}
\end{equation}   
for suitable reservoir operators. Then, provided there is no constraint 
on the rate of control so that condition (\ref{condition}) can be assumed, 
one has to ensure the existence of a gate sequence
generating the required temporal average:
\begin{equation}
\sum_{k=0}^{n_P} \, \Delta t_k \, {\cal P}_k \sigma_\alpha {\cal P}_k =0 \;,
\hspace{2cm}\alpha \in \{x,y,z\}\;. 
\end{equation}
The purely decohering coupling corresponds to ${\cal B}_x={\cal B}_y=0$. 
Actually, binary sequences of the Carr-Purcell type also suffice to decouple
any form of interaction (\ref{arbitrary}) involving at most two error 
generators. A specially relevant case is a Jaynes-Cummings-like 
dissipative coupling with ${\cal B}_z =0$ and {\cite{rasetti}}
\begin{equation}
H_{SB}=\sum_k \, (g_k b_k^\dagger \sigma_- + h.c.) \;\; 
\Rightarrow \;\; {\cal B}_x = \sum_k \, (g_k b_k^\dagger + h.c.)\,,\; 
{\cal B}_y = \sum_k \, (-i g_k b_k^\dagger + h.c.) \;, 
%\;{\cal B}_z =0 \;,
\end{equation}
which can be eliminated, in principle, by a sequence of $\pi$-pulses along the 
$\hat{z}$-axis. A slightly more elaborated sequence is  
necessary to decouple the qubit from the simultaneous action of the three 
error generators. It turns out that it can be derived on the basis of a 
simple group-theoretic argument.  
A more detailed and general formulation of the method is presented  
elsewhere {\cite{no2}}. 
    
\section{CONCLUSIONS}

Our work demonstrates the possibility to modify the evolution of a quantum 
open system by applying external controllable interactions.
From the perspective of quantum information, the analysis suggests a different 
promising direction compared to conventional quantum error-correction 
techniques. The practical usefulness of the proposed approach strongly 
depends, in its present status, on the time scale of the motional processes 
causing relaxation. The effectiveness of analogous schemes under less 
idealized assumptions and in the presence of 
a finite bound on the control rate deserves further investigation, 
together with the possibility of examining decoherence properties within a 
fully quantum mechanical control configuration as recently proposed in 
{\cite{seth1}}.

%%%%%%%%%%%%%%%%%%%%%%%%%%%%%%%%%%%%%%%%%%%%%%%%%%%%%%%%%%%%%%%%%%%%%%%
\acknowledgments
The authors are indebted to Emanuel Knill for many insightful comments.  
This work was supported by ONR, by AFOSR, and by DARPA/ARO under the 
Quantum Information and Computation initiative (QUIC) and the NMR Quantum 
Computing initiative (NMRQC).

\begin{figure}
\caption{ Qubit decoherence as a function of time for an Ohmic environment.
Time is in units of $T^{-1}$ and $\omega_c =100$. High- and low-temperature 
behaviors are depicted, (H) $\omega_c/T=10^{-2}$ and (L) $\omega_c/T=10^2$ 
respectively. }
\end{figure}

\begin{figure}
\caption{ Qubit pulsed decoherence as a function of time for the Ohmic 
high-temperature environ- ment of Fig. 1. A pulse separation 
$\Delta t= \tau_c/10$ has been used and coherence stroboscopically evalua- 
ted using Eqs. (18)-(19). Long-time deviations from unit value arise from 
cumulation of errors in the presence of a small but finite $\Delta t$. }
\end{figure}

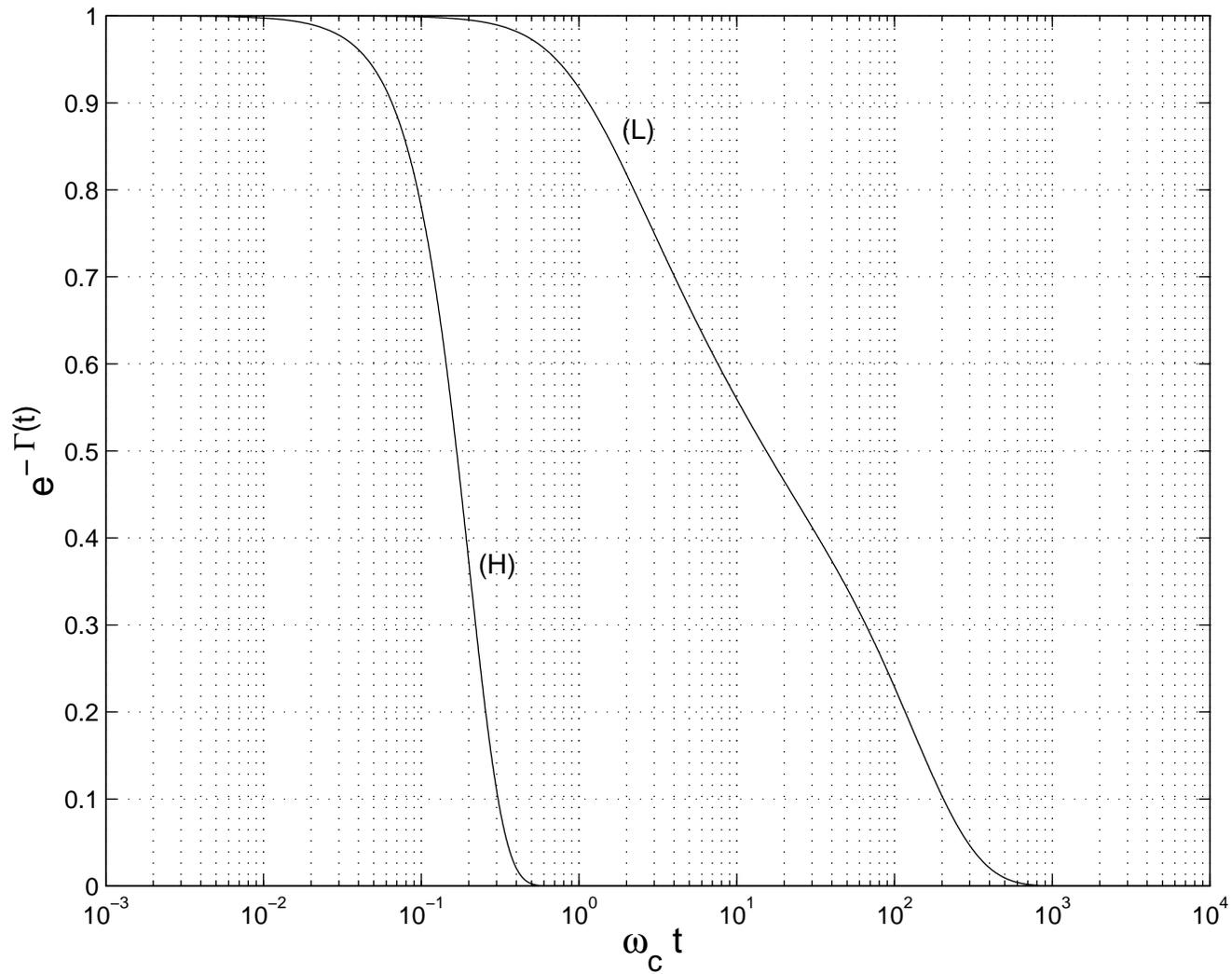
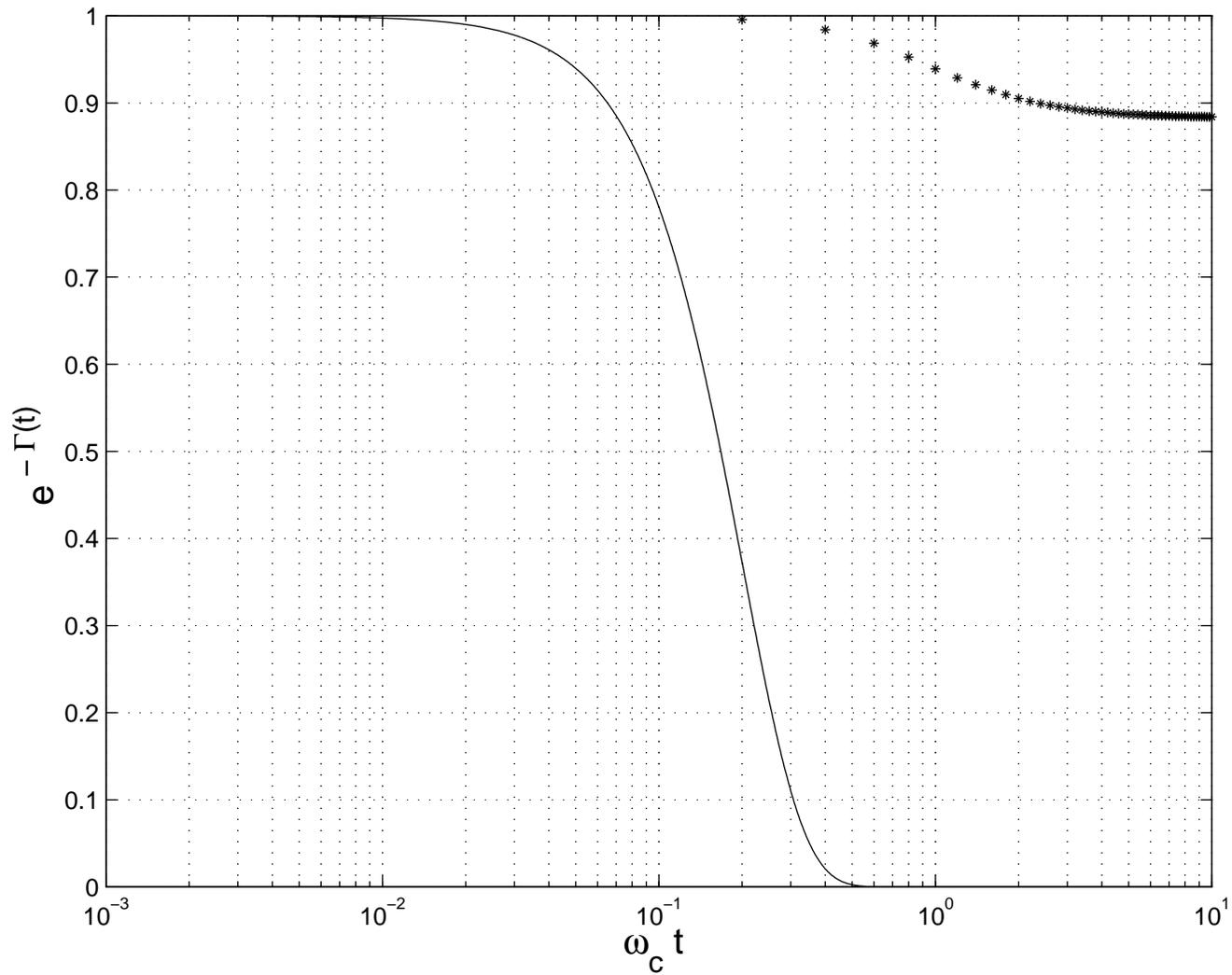
\end{document}